\documentstyle[12pt,fleqn,epsf]{article}
\def\nsection#1{\setcounter{equation}{0}\section{#1}}

\newcommand{\Integer}{\:\mbox{\sf Z} \hspace{-0.82em} \mbox{\sf Z}\,}
\newcommand{\Is}{\mbox{\scriptsize \sf Z}  \! \! \mbox{\scriptsize \sf Z}}
\def\F#1#2#3#4#5{#1\left(\hspace{-1mm}
	 \begin{array}{cc}#5 & #4 \\ #2 & #3 \end{array}
		  \hspace{-1mm}\right)}
\def\C{\rule[0.5pt]{0.2mm}{7pt}{\hspace{-3.6pt}{\rm C}}}
\def\Mult#1#2#3{\left[#1 \atop #2 \right]_{#3}}
\def\Bin#1#2{\left(#1 \atop #2 \right)}
\def\Mults#1#2#3{\left[{\textstyle {#1 \atop #2} } \right]_{#3}}
\def\ib{\,\mbox{i}\,}

\def\e{\mbox{e}}
\def\d#1{\mbox{d}#1}
\def\case#1#2{{\textstyle{#1\over #2}}}
\def\te{\vartheta_1}
\def\la{\lambda}
\def\eps{\epsilon}

\begin{document}

\title{Fermionic solution of the Andrews-Baxter-Forrester model I:
unification of TBA and CTM methods}

\author{S.~Ole Warnaar\thanks{
e-mail: {\tt warnaar@mundoe.maths.mu.oz.au}}
\\
Mathematics Department\\
University of Melbourne\\
Parkville, Victoria 3052\\
Australia}

\date{January, 1995 \\ \hspace{1mm}
\\
Preprint No. 02-95}
\maketitle

\begin{abstract}
The problem of computing the one-dimensional
configuration sums of the ABF model in regime III
is mapped onto the problem of evaluating
the grand-canonical partition function of a gas of
charged particles obeying certain fermionic exclusion rules.
We thus obtain a new {\em fermionic} method to compute the
local height probabilities of the model.
Combined with the original
{\em bosonic} approach of Andrews, Baxter and Forrester, we
obtain a new proof of (some of) Melzer's polynomial identities.
In the infinite limit
these identities yield Rogers--Ramanujan type identities
for the Virasoro characters $\chi_{1,1}^{(r-1,r)}(q)$
as conjectured by the Stony Brook group.
As a result of our working the
corner transfer matrix and thermodynamic Bethe Ansatz
approaches to solvable lattice models are unified. \\
{\bf Key words:} Corner transfer matrices; Thermodynamic
Bethe Ansatz; ABF model; One-dimensional Fermi gas;
Rogers--Ramanujan identities.
\end{abstract}

\nsection{Introduction}
The thermodynamic Bethe Ansatz (TBA) and the
corner transfer matrix (CTM) method  are two of the
most fruitful techniques developed for studying solvable
lattice models.
The TBA approach, originating from the work of Yang and Yang \cite{YY}
and refined and extended in the work of, a.o.,
Takahashi~\cite{Takahashi}, Takahashi and Suzuki~\cite{TS}, and
Bazhanov and Reshetikhin~\cite{BR1,BR2,BR3},
has been applied most successfully
in studying the thermodynamic properties of the energy spectrum of the
row-to-row transfer matrix.
Also, in establishing connections between (off-)critical
solvable lattice models and (perturbed) conformal field theories,
the TBA method has proven to be extremely useful.
The CTM method in turn,
was invented by Baxter (see ref.~\cite{Baxter1} and references
therein) and developed for computing order parameters of solvable
models.

At first sight, both methods seem rather distinct. Different
types of transfer matrices are used for computing
different quantities of interest.
One key-step in both approaches is however similar.\footnote{
Throughout this paper we only refer to models that satisfy the
Yang-Baxter equation with difference property \cite{Baxter1}.}
Given a model system on a finite lattice of size $m$,
one is naturally led to define a ``truncated'' or ``finitized'' system,
which in the limit $m\to\infty$ becomes identical to
the true model of interest.
Let us describe this step in some detail as it turns out to be of
unexpected importance.

In using the Bethe Ansatz technique, one has to solve a coupled set
of non-linear equations, known as the Bethe Ansatz equations (BAE).
These BAE have many solutions corresponding to the different
eigenvalues of the row-to-row transfer matrix.
In solving the BAE numerically for small systemsizes, one
observes that the roots of the BAE occur in so-called strings,
a string being a set of complex  numbers with equal real part but
distinct imaginary parts.
For large $m$, the imaginary
positions of the roots within a string tend to a limit (fixed by the
parameters in the model) and the only
degrees of freedom left in the problem are the positions of the
strings on the real axis, and the ``occupation'' number
of a string of given type.
For finite $m$ the string-picture is violated by:
{\em i}) The roots within a string do not exactly have equal
real part, i.e., the strings are generally slightly curved. {\em ii})
The imaginary parts of the roots within a string deviate
from their fixed $m\to\infty$ position. {\em iii})
A number of roots does not fit the string picture at all, where it is
to be remarked that the number of such roots divided by the total number of
roots tends to zero in the large $m$ limit.\footnote{Recently solutions
to the Bethe equations for the 3-state Potts model have been found
which are neither of string type nor thermodynamically irrelevant.
For the full details on such {\em non-string}
solutions see ref.~\cite{ADM}.}

Based on the above observations, one usually proceeds by
making a {\em string hypothesis}, which amounts to assuming that
all solutions to the BAE consist of sets of strings, each string being
taken from a set of allowed string types.
Of course the string hypothesis, if correct at all, will only describe
the spectrum of the row-to-row transfer matrix in the $m\to\infty$ limit,
and the set of ``reduced'' Bethe equations
obtained after substitution of the
strings makes sense in the thermodynamic limit only.
It is nevertheless tempting to view the reduced equations as those of some
ideal solvable model for which the string hypothesis holds true
even for finite $m$. We call this virtual model the {\em finitized model}.
Of course, the finitized model should not be confused with the
original model on a finite lattice, but should be viewed as
a model on a finite lattice inheriting all simplifying features of
the original model when the lattice size is taken to infinity.

A similar finitization occurs in the CTM calculations.
Provided the corner transfer matrices commute,
Baxter showed that all eigenvalues of the (diagonalized) CTM's
of a solvable model
take a special exponential form, which can be computed explicitly
by considering the ordered low-temperature limit.
Since commutativity of the CTM's holds only in the infinite
lattice limit, we can only diagonalize the CTM's in
this limit.
Let us think of the exponential form of the eigenvalues as being
at the same level as the string hypothesis. Again we can substitute this
form in the relevant finite $m$ equations and think of the equations
thus obtained as those of an ideal solvable model for which the CTM's
commute for finite systemsize. We again call this the {\em finitized
model}.

\vspace*{5mm}

The aim of this paper is to show that given a particular solvable model
the finitized models
obtained from the TBA and from the CTM approach are the same.
To be precise, we show that the finite $m$
TBA and finite $m$ CTM calculations for the finitized
model are mathematically equivalent.  Letting $m$ to infinity,
this establishes the claim that the true (infinite $m$) TBA and CTM methods
are equivalent as well.
We proof the above assertion for what is probably the
most generic example of a solvable lattice model,
the $(r-1)$-state Andrews-Baxter-Forrester (ABF) model \cite{ABF}.
To do so we present a new {\em fermionic} method
for computing the one-dimensional configuration sums of the ABF
model. This method is based on a mapping of the
configuration sums onto the grand-canonical partition function
of a one-dimensional gas of charged fermions.
Up to terminology, this gas is equivalent to the system of
strings and holes describing the TBA.

The remaining sections of this paper are organized as follows.
In section~\ref{sec2} we briefly review the definition of the
ABF model and sketch the {\em bosonic} approach for computing the
local height probabilities as followed  by Andrews {\em et al}.
Then, in section~\ref{sec3}, we present the main ideas of our fermionic
method by computing the simplest possible
configuration sum, using the Fermi gas.
In the subsequent section we discuss Melzer's
polynomial identities~\cite{Melzer1}
and their limiting Rogers--Ramanujan type identities~\cite{KKMM2},
which are
proven by combining the
bosonic and fermionic methods. In section~\ref{sec4} we also
discuss the relevant differences between our approach to
proving Melzer's identities and the original proof given
by Berkovich~\cite{Berkovich}.
In section~\ref{sec5}, we discuss the equivalence between CTM and TBA
as follows from the working of section~\ref{sec3}, and we end with a
summary and discussion in section~\ref{sec6}.

In this paper we have limited ourselves to the fermionic evaluation
of just a single configuration sum.
The computation of all
configuration sums will be the content of a subsequent paper,
referred to as part II.
In this part II, we show that two
additional boundary particles with fractional charges
have to be introduced on top of the Fermi gas
to treat the most general configuration sums. A brief explanation
of the origin of these boundary particles is given in section~\ref{map}

Before we start reviewing the aspects of the ABF model relevant to this
paper, several final introductory remarks
have to be made to put some of the ideas and results of this paper in
proper context. First of all, we have to mention the remarkable
series of papers written by the Stony Brook group
[11,13-15].
Based on a study of the completeness of states
of the Bethe Ansatz solution
for the 3-state Potts model
\cite{KM,DKMM}, large classes of new expressions for conformal characters
where conjectured \cite{KKMM1,KKMM2}. Because
of exclusion rules labelling the Bethe states, these character expressions
were called fermionic. Combining the fermionic forms with
the well-known Rocha-Caridi type bosonic character expressions,
led to conjectures of many new identities similar to those of Rogers and
Ramanujan.
Second,
In ref.~\cite{JMO} Jimbo {\em et al.} pointed out that computing
one-dimensional configuration sums using the recursion method of ABF
leads to bosonic character expressions.
This, together with the fermionic forms originating from the Bethe
Ansatz approach indicates a deep connection between (T)BA and CTM
methods.
Signs of an even stronger connection were found by Melzer \cite{Melzer1},
who conjectured that finitizations of the fermionic character expressions,
originating from TBA for the finitized ABF model, equal finitizations
of the bosonic character expressions originating from CTM
for the finitized ABF model.
It has especially been this last result and its many generalizations to
other solvable models [18-26],
that has motivated our attempt to
unify TBA and CTM approaches to solvable models.

\nsection{The ABF model}\label{sec2}
This section serves as a brief introduction to the ABF model,
and, apart from the model definition, reviews part of the
calculation of the order parameters for regime III
as carried out in ref.~\cite{ABF}.

\subsection{Definition of the model}
The ABF model is a restricted solid-on-solid (RSOS) model
defined on the square lattice $\cal L$.
Each site $i$ of $\cal L$ carries a spin or height variable $\sigma_i$,
which can take any of the  values $1,2,\ldots,r-1$.
A restriction is imposed on each pair of neighbouring sites $i$ and $j$
by $|\sigma_i-\sigma_j|$=1.
The Boltzmann weight of a configuration of heights on the lattice
is computed as the product over elementary face-weights,
with nonzero faces parametrized as
\begin{eqnarray}
\F{W}{a}{a \pm 1}{a}{a \mp 1}&=&
\frac{\te(\la-u)}{\te(\la)}
\nonumber \\  & & \nonumber \\
\F{W}{a \pm 1}{a}{a \mp 1}{a}&=&
\left(\frac{\te((a+1)\la) \te((a-1)\la)}{\te^2(a\la)} \right)^{1/2}
\frac{\te(u)}{\te(\la)}  \label{weights}
\\  & & \nonumber \\
\F{W}{a \pm 1}{a}{a \pm 1}{a}&=&
\frac{\te(a \la \pm u)}{\te(a\la)} .
\nonumber
\end{eqnarray}
Here we have used the elliptic theta function
with argument $u$ and arbitrary but
fixed nome  $p$, $|p|<1$,
\begin{equation}
\te(u) = 2 \, p^{1/8} \sin u \prod_{n=1}^{\infty}
\left(1-2p^n \cos 2 u + p^{2n}\right)\left(1-p^n\right).
\end{equation}
The variable $u$ in (\ref{weights}) is the usual spectral parameter,
and $\la$ is the crossing parameter, fixed by the number of states
of the model as
\begin{equation}
\la=\frac{\pi}{r}.
\label{crossing}
\end{equation}
In ref.~\cite{ABF} the weights $W$ were shown to solve the
Yang-Baxter equation \cite{Baxter1}, and hence quantities like
the free energy and order parameters can be computed exactly.
In their calculations, ABF distinguished four different
physical regimes, but here
we restrict ourselves to
regime III, given by
\begin{equation}
\mbox{regime III:} \qquad 0<u<\la, \qquad 0<p<1.
\end{equation}

\subsection{Computation of the local height probabilities}
As a first step in calculating the order parameters of the
ABF model
(see ref.~\cite{Huse} for the actual definitions), one has to compute
the so-called local height probabilities $P^{bc}(a)$.
Here $P^{bc}(a)$ is the probability that a spin of the (infinite) lattice
takes the value $a$ provided that the model is in  a phase
indexed by $b$ and $c$.
In their computation of
$P^{bc}(a)$ ABF start with  a finite lattice $\cal L$ with
size measured by $m$,
and consider  the probability $P_m^{bc}(a)$ that the center
spin of $\cal L$ takes the value $a$, provided the boundary spins
on $\cal L$ are fixed according to a groundstate labelled by $b,c$.
For regime III
there are $2r-4$ antiferromagnetic groundstates, corresponding to
all spins on  the even sublattice taking the value $b$ and
all spins on the odd sublattice taking the values $c=b\pm 1$,
$b=(3\mp 1)/2,(5\mp 1)/2, \ldots, (2r-3\mp 1)/2$.
We can thus write
\begin{equation}
P^{bc}(a) = \lim_{m\to \infty} P_m^{bc}(a),
\label{Pm}
\end{equation}
with $b,c$ taking any of the above groundstate values.
At the same time ABF showed, using the corner transfer matrix method,
that
\begin{equation}
P^{bc}(a) = \lim_{m\to\infty} \frac{E(x^a,x^r) X_m(a,b,c;x^2)}
{\sum_{a=1}^{r-1} E(x^a,x^r) X_m(a,b,c;x^2)} \: ,
\label{Pbca}
\end{equation}
with the {\em one-dimensional configuration sums} defined as
\begin{equation}
X_m(a,b,c;q) = \sum_{
\begin{array}{c} \scriptstyle
\sigma_1,\ldots,\sigma_{m-1}=1\\
\scriptstyle
|\sigma_{j+1}-\sigma_j|=1
\end{array}}^{r-1}
q^{\sum_{k=1}^m
k|\sigma_{k+1}-\sigma_{k-1}|/4},
\qquad  \sigma_0=a, \; \sigma_m=b, \; \sigma_{m+1}=c.
\label{confsums}
\end{equation}
The nonstandard labelling of the $\sigma$'s from 0 to $m+1$
is chosen for later convenience.
The variable $x$ in (\ref{Pbca})  relates to the nome $p$
of the Boltzmann weights as
\begin{equation}
x=\e^{-4 \pi/(r\eps)}, \qquad \mbox{ with } \quad  p=\e^{-\eps}.
\end{equation}
The function $E$ in (\ref{Pbca}) is the elliptic function
\begin{equation}
E(z,p) = \prod_{n=1}^{\infty} (1-p^{n-1} z)(1-p^n z^{-1})(1-p^n),
\end{equation}
for all $z,p \in \: \C$, $|p|<1$.

We remark here that in deriving (\ref{Pbca}), one has to assume
commuting CTM's, and hence that the size of the lattice
is infinite. Therefore, the parameter $m$ in (\ref{Pbca}) is not
the  same as the systemsize $m$ in (\ref{Pm}).
As described in the introduction, we now define our finitized model
by identifying the $m$'s in (\ref{Pbca}) and (\ref{Pm}).
In other words, we define the finitized model through $P_m^{(bc)}(a)$ as
\begin{equation}
P_m^{bc}(a) =
\frac{E(x^a,x^r) X_m(a,b,c;x^2)}
{\sum_{a=1}^{r-1} E(x^a,x^r) X_m(a,b,c;x^2)} \: .
\end{equation}

The problem now is to perform the sum in (\ref{confsums}) to
obtain manageable closed form expressions for the configuration
sums $X_m(a,b,c;q)$.
In the remainder of this paper we will give two approaches to the
solutions of this problem. The first is the method originally
employed by Andrews, Baxter and Forrester \cite{ABF}.
It results in expressions for the configuration sums as the difference
of two finite $q$-series, and hence, adopting the terminology
introduced by the Stony Brook group \cite{KKMM1,KKMM2},
we call this method bosonic.
The second method is new, and amounts to interpreting the sum in
(\ref{confsums}) as the grand-canonical partition function of a
one-dimensional gas of charged particles, obeying certain
exclusion rules.  This time the resulting expressions involve just
a single finite $q$-series.
Again following ref.~\cite{KKMM1,KKMM2}, we call the second method
fermionic.\footnote{We remind the reader that Baxter's original method of
computing $P^{bc}(a)$ for the hard hexagon model also
led to infinite fermionic $q$-series \cite{Baxter2}.
Our method can thus be seen
as a generalization of the working of ref.~\cite{Baxter2}.}

\subsection{Bosonic evaluation of the configuration sums}
Let us begin with a brief reminder of the original
bosonic approach of Andrews, Baxter and Forrester, as
detailed in ref.~\cite{ABF}.

First, notice that the $X_m(a,b,c;q)$ can be (re)defined
in terms of the following recurrence relations:
\begin{equation}
X_m(a,b,b\pm 1) = X_{m-1}(a,b\pm q,b) + q^{m/2}
X_{m-1}(a,b\mp 1,b), \qquad  3
\leq 2 b\pm 1 \leq 2 r-3,
\label{recrel}
\end{equation}
subject to the initial and boundary conditions
\begin{eqnarray}
\lefteqn{
X_0(a,b,c) = \delta_{a,b} \, \delta_{c,b \pm 1},} \nonumber \\
& & \nonumber \\
\lefteqn{
X_m(a,0,1)=X_m(a,r,r-1)=0.}
\label{iandb}
\end{eqnarray}
Then, introducing the Gaussian or $q$-binomials as
\begin{equation}
\Mult{N}{m}{q} =  \left\{
\begin{array}{ll}
\displaystyle \frac{(q)_N}{(q)_m (q)_{N-m}} \quad & 0\leq m \leq N \\
& \\
0 & \mbox{otherwise},
\end{array} \right.
\end{equation}
with $(q)_m = \prod_{k=1}^m (1-q^k)$ for $m>0$ and $(q)_0=1$ ,
we have the following theorem \cite{ABF}:

\vspace*{2mm}
{\bf Theorem:}
For $m\geq 0$, $1 \leq a,b,c \leq r-1$, $c=b\pm 1$, $m+a-b \in
\Integer_{\geq 0}$, the one-dimensional configuration sums read
\begin{eqnarray}
X_m(a,b,c) &=& q^{(a-b)(a-c)/4} \sum_{j=-\infty}^{\infty} \left\{
q^{r(r-1)j^2+[r(b+c-1)/2-(r-1)a]j}
\Mult{m}{\frac{1}{2}(m+a-b)-rj}{q} \right.
\nonumber \\ & &  \nonumber \\
& & \left. \qquad \quad
-q^{r(r-1)j^2+[r(b+c-1)/2+(r-1)a]j+a(b+c-1)/2}
\Mult{m}{\frac{1}{2}(m-a-b)-rj}{q} \right\}.
\label{the1}
\end{eqnarray}
For the proof of this result we refer the reader to ref.~\cite{ABF}.
Let us however remark that it follows in straightforward
manner substituting (\ref{the1}) into (\ref{recrel})
and using the elementary recurrences \cite{Andrews}
\begin{eqnarray}
\Mult{N}{m}{q}&=& \Mult{N-1}{m}{q}+ q^{N-m} \Mult{N-1}{m-1}{q}
\nonumber \\
& & \nonumber \\
&=& \Mult{N-1}{m-1}{q}+ q^{m} \Mult{N-1}{m}{q} .
\label{qdecomp}
\end{eqnarray}
Also the conditions (\ref{iandb}) are readily checked for the expression
(\ref{the1}).

Before concluding this section we note that the $X_m(a,b,c;q)$
can be viewed as {\em finitizations} of the Rocha-Caridi
expressions for the (normalized) Virasoro characters
$\chi_{(b+c-1)/2,a}^{(r-1,r)}(q)$ \cite{Rocha},
\begin{eqnarray}
\lefteqn{
\lim_{m\to\infty} q^{-(a-b)(a-c)/4} X_m(a,b,c;q) } \nonumber \\
& & \nonumber \\
&=&
\frac{1}{(q)_{\infty}}
\sum_{j=-\infty}^{\infty} \left\{
q^{r(r-1)j^2+[r(b+c-1)/2-(r-1)a]j}
-q^{r(r-1)j^2+[r(b+c-1)/2+(r-1)a]j+a(b+c-1)/2}
\right\} \\
& & \nonumber \\
& \equiv & \chi_{(b+c-1)/2,a}^{(r,r-1)}(q). \nonumber
\end{eqnarray}

\nsection{Fermionic evaluation of the configuration sums}\label{sec3}
We now come to the main part of this paper, the
alternative method to evaluate the sum (\ref{confsums}).
As mentioned earlier, we can interpret this sum as the grand-canonical
partition function of a one-dimensional gas of charged particles.
Once this identification has been made, the problem of course
remains to actually compute this partition function.
For pedagogical reasons this is in fact what we do first, that is,
we start by defining our system of particles and compute its
generating function. Only afterwards we then perform the
actual mapping from the one-dimensional configuration sums
onto the Fermi gas partition function.
To keep things as simple as possible we restrict ourselves to the
evaluation of $X_m(1,1,2)$. The case of arbitrary $X_m(a,b,c)$,
which involves the additional concept of
boundary particles with fractional
charge, will be
treated in part II of this paper.

\subsection{A one-dimensional Fermi gas}\label{secCG}
\begin{figure}[hbt]
\centerline{\epsffile{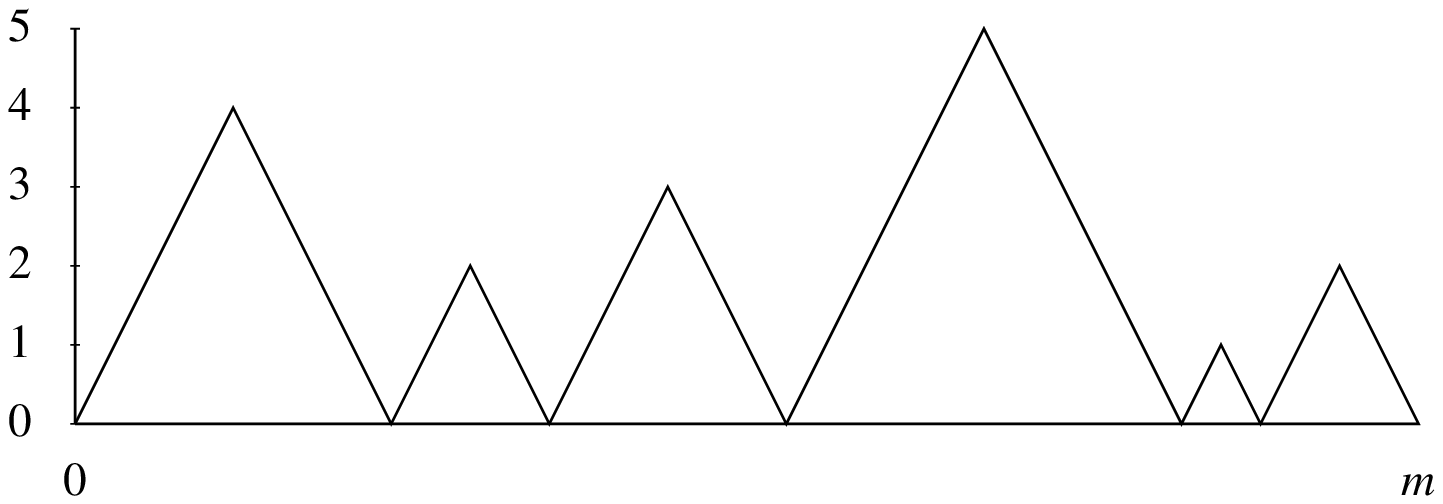}}
\caption{A possible charge configuration of the Coulomb gas for $r\geq 7$.}
\label{fig1}
\end{figure}

We consider a system of charged particles occupying
a one-dimensional lattice of $m+1$ sites, with sites
labelled from $0$ to $m$, $m$ being even.
The number of particles with charge $j$ is
$n_j$ ($j=1,\ldots,r-2$) and a particle of charge $j$
has diameter $2j$.
The lattice is completely occupied with particles, yielding
the completeness relation
\begin{equation}
2 \sum_{j=1}^{r-2} j \, n_j = m.
\label{completeness}
\end{equation}
A typical configuration of particles is shown in figure~\ref{fig1},
where we have drawn a particle of charge $j$
as a triangle with {\em height} $j$.

\begin{figure}[hbt]
\centerline{\epsffile{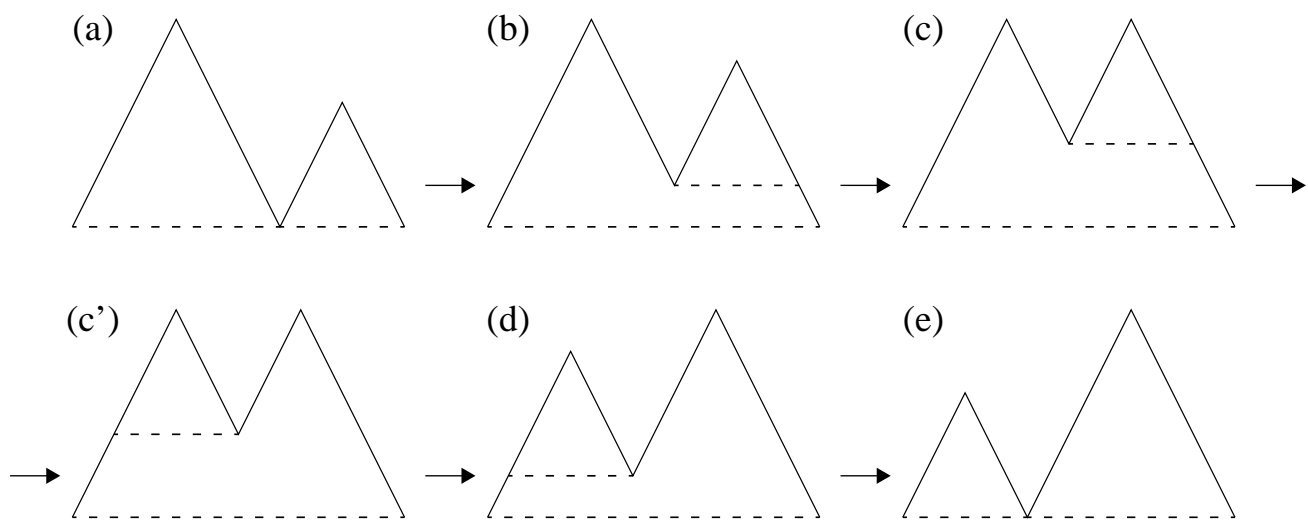}}
\caption{The movement of a particle of charge $k$ through a particle
of charge $j>k$.
According to rule R3 the configurations in figure c and c' are to
be identified.
The dashed lines are the baselines as described on page~8.
We note that the procedure for drawing the baselines in fact
rules out possibility c' thus avoiding overcounting
of configurations.}
\label{fig2}
\end{figure}

To describe the allowed motion of the particles, we have the
following  set of rules:
\begin{description}
\item[R1]
Hard-core repulsion between particles of equal charge.
Therefore, given an ``initial'' configuration, the order of the particles
with charge $j$ remains fixed. For obvious reasons we can therefore
think of the particles as being fermions.
\item[R2]
Two particles with charge $j$ and $k$, $j \neq k$ can penetrate each
other, and eventually, exchange position. \label{pR2}
An example of the motion of a particle with charge $k$ through a particle
with charge $j>k$ is shown in figure~\ref{fig2}.
We note that the diameter of the {\em charge complex} in this figure
remains $2(j+k)=2j + 2k$ throughout the motion and hence that we
obey charge conservation. We also note that both the
position  and the number of peaks of the complex remain fixed.

\item[R3]
We identify configurations that lead to the same profile
as being the same. Hence the charge configurations c and c' in
figure~\ref{fig2} are identical. (Although R1 is a special
case of R3, we wish to state it as a separate rule.)
\end{description}

\begin{figure}[hbt]
\centerline{\epsffile{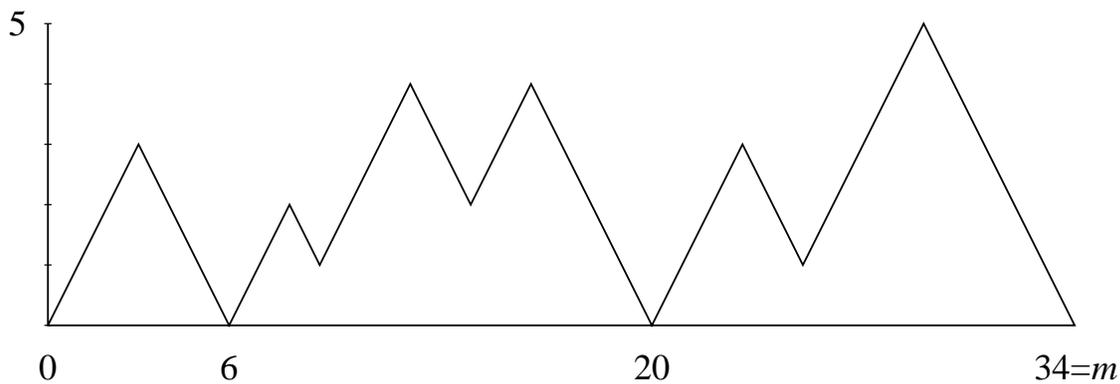}}
\caption{A typical charge configuration in which
some of the particles have formed charge complexes.}
\label{fig3}
\end{figure}

With the above rules it is clear that a configuration as drawn in
figure~\ref{fig1} is not generic, since all particles are properly
separated and no charge complexes occur.
A more general situation, which can be obtained from
the configuration of figure~\ref{fig1} by moving around the
particles following the rules of motion R1-R3,
is shown in figure~\ref{fig3}.

\begin{figure}[hbt]
\centerline{\epsffile{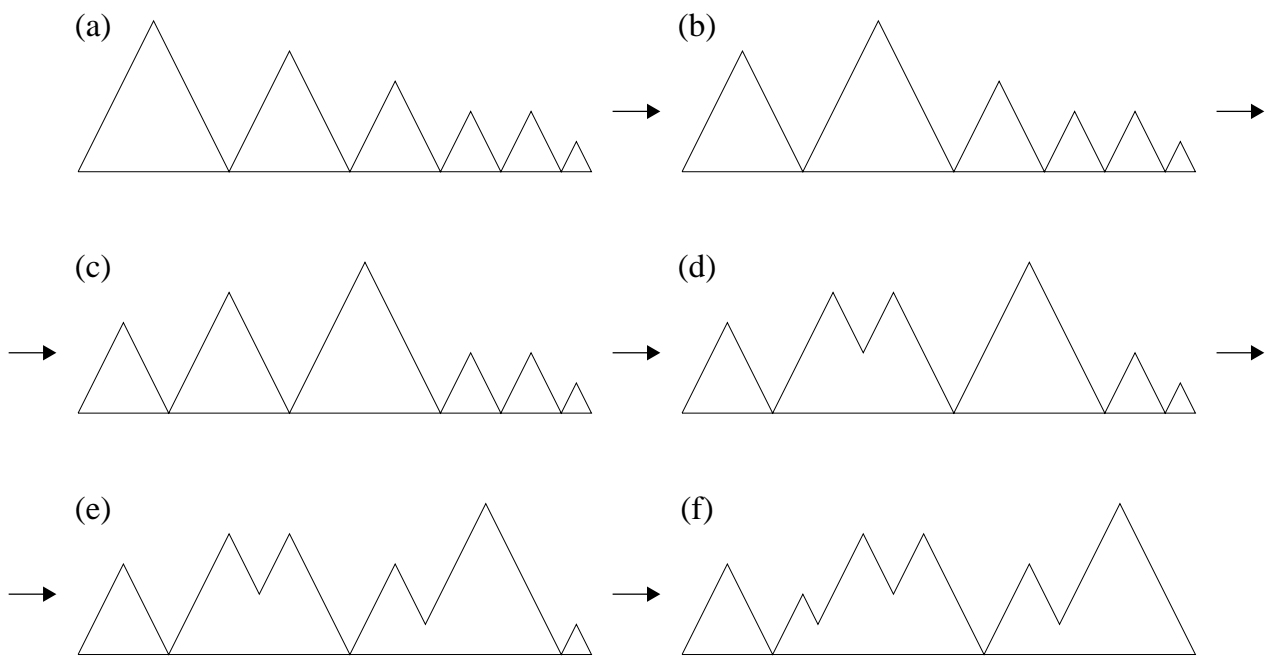}}
\caption{The construction of the charge configuration of
figure~3 out of a minimal configuration
by carrying out the steps described under M1-M6.}
\label{fig4}
\end{figure}

\begin{figure}[hbt]
\centerline{\epsffile{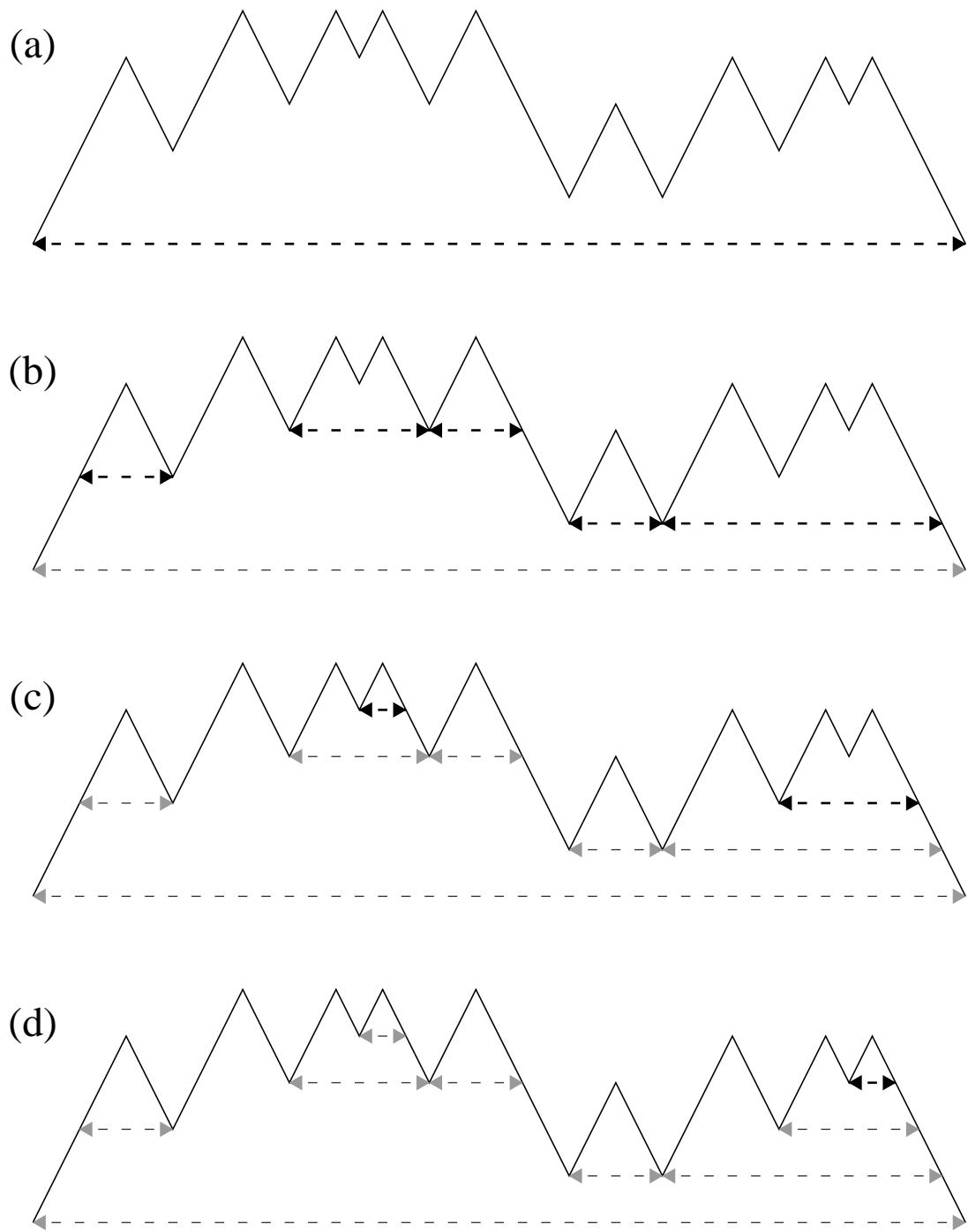}}
\caption{The drawing of baselines in a charge complex
to determine its particle content.
The small arrows indicate how to divide the $j$-th
baseline into zeroth baselines of smaller complexes.
(a) Drawing of the zeroth baseline.
(b) Drawing of the first baseline.
(c) Drawing of the second baseline.
(d) Drawing of the third baseline.}
\label{fig5}
\end{figure}

We now come to a crucial observation about our Fermi
gas.
Fix the number of particles of given charge, i.e., fix
$n_1,n_2,\ldots,n_{r-2}$, and call the configuration with
all particles being separated, and all particles of charge $j$
being positioned to the left of all particles of charge $k<j$,
a {\em minimal configuration}.\footnote{This terminology, suggested
to the author by O.~Foda,
stems from a partition theoretic interpretation
of minimal configurations.}
Then all other configurations with the same particle content,
i.e., with the same $n_1,n_2,\ldots,n_{r-2}$ can be obtained by
the following motion starting from the minimal configuration:
\begin{description}
\item[M1]
Keep the particles with charge $r-2$ fixed.
\item[M2]
Move the left-most particle of charge $r-3$, denoted
$p_{r-3,1}$ to the left through
some (and possibly all) of the particles of charge $r-2$.
It can either end up as a charge complex
or as a separate particle.
\item[M3]
Move the second-left-most particle of charge $r-3$
(denoted $p_{r-3,2}$) to the left.
Of course, this time the movement to the left is bounded by the
position of particle $p_{r-3,1}$ due to the fermionic
exclusion rule R1.
\item[M4]
Continue to move all particles $p_{r-3,n}$, $n=1,2,\ldots,n_{r-3}$
to the left, one at the time and $p_{r-3,n+1}$ after $p_{r-3,n}$.
Each time the motion of particle $p_{r-3,n+1}$ is bounded to the left
by the position of particle $p_{r-3,n}$.
\item[M5]
After having moved the particles of charge $r-3$,
start a similar procedure for the particles of charge $r-4$.
The left-most particle $p_{r-4,1}$ first and right-most particle
$p_{r-4,n_{r-4}}$ last.

\item[M6]
Continue this process till all particles have moved to the left,
always obeying the order: $p_{j,n}$ before $p_{j,n'}$ for $n<n'$
and $p_{j,n}$ before $p_{k,n'}$ for $j>k$.
Each time the particle $p_{j,n+1}$ is bounded to the
left by $p_{j,n}$, with $p_{j,1}$ bounded by the
origin of the lattice.
\end{description}
In figure~\ref{fig4} we have shown how to obtain the configuration
of figure~\ref{fig3} by consecutively moving
the particles of the corresponding
minimal configuration leftwards.
To proof that we indeed can obtain all allowed configurations by
starting from a minimal configuration carrying out the above
steps, we first note that given a configuration, e.g., that of
figure~\ref{fig3}, we can uniquely determine its particle content
in the following way:
\begin{itemize}
\item
First locate all charge complexes, regarding a separate particle
as a complex (though trivial) as well.
This amounts to marking all coordinates on the lattice that
have zero height.
In the case of the configuration of figure~\ref{fig3},
the first complex ranges from 0 to 6, the second from 6
to 22 and the third from 22 to $34=m$.
\item
For each separate complex we now determine its particle content as follows:
\begin{itemize}
\item
Draw a dashed line along the lattice from the origin (left-most point)
of the complex to its endpoint (right-most point).
We call this the zeroth baseline of the complex.

The result of this trivial step is shown in figure~\ref{fig5}a
for a typical charge complex.

\item
Start from the highest point of the complex.
If there is more than one highest point, start from the left-most
highest point.
Move down to the right of the peak along the contour of the complex till
its endpoint. (The point were the next complex starts.)
If a local minimum is reached, i.e., the
contour of the complex starts going up again, we draw a dashed line
from this local minimum to the right, until we cross the contour
of the complex. At that point we move further down along the
contour. If another minimum occurs we repeat the above, et cetera.

We do exactly the same, now starting from the same left-most highest point
but moving down to the left. If a local minimum is reached we draw a dashed
line to the left and continue our movement down when the dashed line
intersects the contour of the complex.

The result of the above procedure is shown in figure~\ref{fig5}b.
The total of dashed line segments drawn in this step is called
the first baseline.

\item
We now view the first baseline as the zeroth baseline of smaller complexes.
Let there be $L$ such smaller complexes. We then
divide the first baseline into $L$ pieces by marking the begin- and
endpoint of these complexes by arrows
pointing to the left and right, respectively, see figure~\ref{fig5}b.

Either these complexes consist of a single particle (complexes with
one maximum and consequently no local minima), or they are again
complexes in the true sense of the word. In the latter case, we
repeat the procedure of drawing dashed lines, again starting
from the left-most highest point.

The result of this third round of drawing dashed lines is shown in
figure~\ref{fig5}c.
The total of dashed line drawn in this step is called
the second baseline.
Again this baseline is divided into pieces, marking yet smaller
complexes.

\item
We repeat the above procedure of drawing and cutting up higher
order baselines iteratively.
That is, we view the $j$-th baseline as the zeroth baseline of yet
smaller objects, we draw arrows accordingly and, if some of these
smaller complexes have more than one maximum, we start drawing the
$(j+1)$-th baseline.

Since the highest point of the complex is always less than $r-1$,
the iteration terminates within $r-2$ steps (including the
drawing of the zeroth baseline).
In the case of figure~\ref{fig5}, we are done in four iterative steps,
figure~\ref{fig5}d being the final result.

\item
Given the complex with all the baselines we can read of its
particle content straight away.
First of all, according to R2,
each peak corresponds to a particle. To determine its
charge we move down vertically from its peak
till me meet the zeroth baseline
relative to this particle, i.e., we move down till we first
intersect a baseline.
The height of the peak minus the height of the baseline
at the intersection point is its charge.
In the example of figure~\ref{fig5}d, we thus get
$(n_1,n_2,\ldots) = (2,5,1,0,1,0,0,\ldots)$.
\end{itemize}

\item
Clearly we now determine the content of a given configuration as the
sum of the particle content of all its complexes.
\end{itemize}
We note that in solving the problem of determining the particle
content of a configuration we have actually localized
each individual particle.
Hence we can start moving the particles to the right to obtain the
minimal configuration. We perform this motion in exactly the opposite order
as described under M1-M6.
Since our rules of motion are reversible (left-right invariant)
we have consequently proven our assertion that each configuration
can be obtained from a minimal configuration following the previously
described rules.

\begin{figure}[hbt]
\centerline{\epsffile{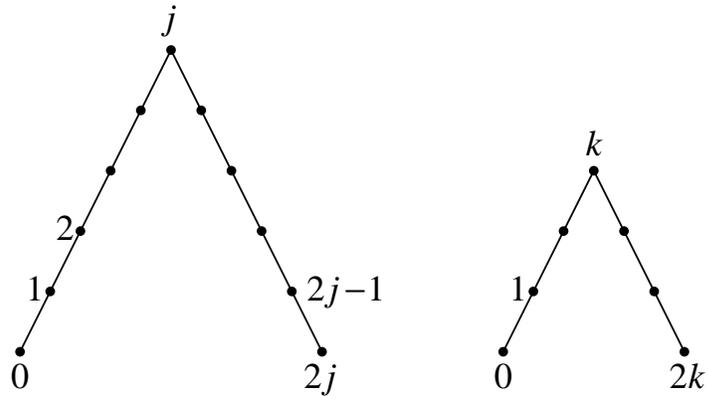}}
\caption{Particles of charge $j$ and $k$
with labelling of begin-, end- and
interior points.}
\label{fig6}
\end{figure}

A final ingredient needed in the Fermi gas is the actual Boltzmann weight
of a given charge configuration. Given this we can try to compute the
partition function.
Before we describe this missing piece of information,
let us first compute
the total number of configurations keeping the particle content
fixed, as this naturally
leads us to introduce some final notation needed.
Consider the particle of charge $j$ as shown in figure~\ref{fig6}.
We denote the left-most point of the particle its {\em origin}, and its
right-most point its {\em endpoint}. All points $1,2,\ldots,2j-1$
will be referred to as {\em interior points} of the particle.
Let us now ask the question in how many ways $I(j,k)$ this particle
and the particle with charge $k<j$ shown in the same figure,
can form a charge complex.
Clearly, (see also figure~\ref{fig2}), we can place the origin
of the particle with charge $k$
at the following interior points of the particle
of charge $j$:
\begin{equation}
2j-1,2j,\ldots,j+k.
\end{equation}
Similarly, we can place the endpoint of the particle at
the following interior points:
\begin{equation}
1,2,\ldots,j-k.
\end{equation}
If we now remark that,
according to rule R3,
placing the origin at $j+k$ gives
the same configuration as placing the endpoint at $j-k$,
(see figure~\ref{fig2}c and c')
we have a total of
\begin{equation}
I(j,k)=2j-2k-1
\label{complex}
\end{equation}
ways to form a complex of charge $j+k$.

This prepares us to answer the combinatorial question raised
above.
The number of ways to move the particles with charge $r-3$ to the
left, starting from the minimal configuration, is
\begin{equation}
\sum_{k_1=0}^{2n_{r-2}} \sum_{k_2=0}^{k_1} \ldots
\sum_{k_{n_{r-3}}=0}^{k_{n_{r-3}-1}} 1 =
\Bin{2 n_{r-2} + n_{r-3}}{n_{r-3}},
\label{move1}
\end{equation}
since $I(r-3,r-4)=1$, and since we can position a
particle not only in the interior of a larger particle as a
charge complex, but also
in between two larger particles as a separate unbound particle.

Similarly,
the number of ways to move the particles with charge $r-4$ to the
left is
\begin{equation}
\sum_{k_1=0}^{4n_{r-2}+2n_{r-3}} \sum_{k_2=0}^{k_1} \ldots
\sum_{k_{n_{r-4}}=0}^{k_{n_{r-4}-1}} 1 =
\Bin{4 n_{r-2} + 2 n_{r-3} + n_{r-4}}{n_{r-4}},
\end{equation}
since $I(r-2,r-4)=3$ and $I(r-3,r-4)=1$.
In the general case with particles of charge $j$ we obtain
\begin{eqnarray}
\lefteqn{
\sum_{k_1=0}^{2(r-2-j)n_{r-2}+ \ldots + 4n_{j+2}+2n_{j+1}}
\sum_{k_2=0}^{k_1} \ldots
\sum_{k_{n_j}=0}^{k_{n_j-1}} 1  } \nonumber \\
& & \nonumber \\
&&  \qquad \qquad \qquad \qquad  =
\Bin{2(r-2-j)n_{r-2} + \ldots + 4 n_{j+2} + 2n_{j+1}+n_j}{n_j},
\label{genj}
\end{eqnarray}
as $I(r-2,j)=2(r-2-j)-1,\ldots,I(j+1,j)=1$.
Of course the above calculation is rather clumsy, and we could have
written the binomial answers straight away. As we later need the
$q$-analogues of (\ref{move1})-(\ref{genj})
we nevertheless thought it instructive to
present the result in the above written form.

Collecting the above results we get for the total number
of configurations $Z$ with fixed particle content $n_1,n_2 \ldots
n_{r-2}$,
\begin{equation}
Z(n_1,n_2,\ldots,n_{r-2})=
\prod_{j=1}^{r-3}
\Bin{2\sum_{k=0}^{r-3-j} (r-2-j-k) n_{r-2-k} + n_j}{n_j}.
\end{equation}
In turn summing over different particle content, computing the number
of configurations $\Xi_m$ in a grand-canonical setting, we get
\begin{equation}
\Xi_m = \left.
\sum_{n_1,n_2,\ldots,n_{r-2}\geq 0}
\right.^{\!\!\!\!\!\!\!\!\!\!\!\!\!\!\!\!\!'}
\qquad Z(n_1,n_2,\ldots,n_{r-2}),
\end{equation}
where the prime over the sum indicates the restriction
(\ref{completeness}), i.e., we sum over particle numbers keeping
the size of the system fixed.

This last result suggests in fact to eliminate one degree of
freedom. To do so we introduce the variables $m_j$ as
the number of antiparticles of charge $-j$,
\begin{equation}
m_j=2 n_{j+1} + 4 n_{j+2} + \ldots + 2(r-2-j) n_{r-2}, \qquad
j=1,\ldots,r-2.
\label{antip}
\end{equation}
{}From this we see that antiparticles of charge $-(r-2)$ are absent,
and hence we can eliminate the dependence on the particles of
charge $r-2$.
Using the vector notation $\vec{m}=(m_1,m_2,\ldots,m_{r-3})^T$,
$\vec{n}=(n_1,n_2,\ldots,n_{r-3})^T$ and $(\vec{\e}_1)_j=\delta_{1,j}$,
and denoting the incidence matrix of the
A$_{r-3}$ Dynkin diagram as $\cal I$,
we find
\begin{equation}
\vec{m} + \vec{n} = \frac{1}{2} \left(
{\cal I} \: \vec{m} + m  \: \vec{\e}_1 \right),
\label{constraint}
\end{equation}
and
\begin{equation}
\Xi_m=
\left. \sum \right.^{'} \; \prod_{j=1}^{r-3}
\Bin{m_j+n_j}{m_j}.
\label{Fm}
\end{equation}
Here the sum is over all (integer) solutions to the
constraint equation (\ref{constraint}), where the prime denotes
the restriction to even antiparticle numbers $m_j$,
as follows from (\ref{antip}).
Equation (\ref{constraint}), relating the variables $n_j$ to the
new variables $m_j$, originates from the work on
fermionic sums by Berkovich~\cite{Berkovich}.
As we will see in section~\ref{sec5}, it also
appears in the context of TBA.

\subsection{Computation of the partition function}
We now finally come to the actual definition of the Boltzmann weight
of a charge configuration, and to the calculation of the
grand-canonical partition function.

Let us return to the charge configuration of figure~\ref{fig3}.
Clearly, at each (integer) point $x$ $(x=1,\ldots,m-1)$
of the lattice there are four
possibilities for the contour of the
configuration.\footnote{
To be precise we should of course say ``contour in a small
neighbourhood of position $x$.''}
We either have a straight line going up or down,
or there is a cusp corresponding to a (local) minimum or maximum.
We now assign an energy $E(x)$ to each site $x$ of the lattice by
\begin{equation}
-\beta E(x) = \left\{
\begin{array}{ll}
\frac{1}{2} x \log q
\qquad & \mbox{if the contour at position $x$
is a straight line} \\
0 & \mbox{if the contour at position $x$ is a cusp.}
\end{array}\right.
\label{energy}
\end{equation}
Clearly, with this definition the groundstate of the model
is given by the state with only particles of charge 1.
For given, fixed  particle content, the energy is minimized by
ordering the particles to form a minimal configuration.

The remainder of this subsection
is devoted to the computation of the grand-canonical partition function
of our Fermi gas,
\begin{equation}
\Xi_m(q) = \sum_{\mbox{\scriptsize particle content}}
Z(n_1,\ldots,n_{r-2};q) ,
\end{equation}
with the partition function of fixed particle content given by
\begin{equation}
Z(n_1,\ldots,n_{r-2};q)
= \sum_{\mbox{\scriptsize charge configurations}}
\e^{\displaystyle -\beta \sum_{k=1}^{m-1} E(k)}.
\end{equation}
To compute $Z$ we follow a procedure
similar to that employed in the section~\ref{secCG}.
First we consider the minimal configuration, and from that we
obtain all other configurations with equal content by carrying
out the steps M1-M6.
The only extra input will be that changing the position of a particle
changes the energy of a configuration.

To compute the energy of a minimal configuration
we proceed as follows.
The energy of a particle with charge $j$, with its origin
at position $x$ is
\begin{equation}
-\beta E(j,x)=
\frac{1}{2} \log q \sum_{
k=1, \; k\neq j
}^{2j-1} (k+x)  = (j+x)(j-1) \log q
\end{equation}
Hence, for the minimal configuration with content $n_1,\ldots,n_{r-2}$
we compute
\begin{eqnarray}
-\beta E_{\mbox{\scriptsize min}}
&=& \log q \sum_{j=1}^{r-2} \sum_{\ell=1}^{n_j}
E\left(j,2j(\ell-1) + 2\sum_{k=j+1}^{r-2} k \, n_k \right)
\nonumber \\
& & \nonumber \\
&=& \log q
\sum_{j=1}^{r-2} (j-1) n_j \left( j \, n_j +2 \sum_{k=j+1}^{r-2} k \, n_k
\right) \\
& & \nonumber \\
&=& \log q \sum_{j=1}^{r-3}  \sum_{k=1}^{r-3} n_{j+1} \,
A_{j,k} \, n_{k+1}, \nonumber
\end{eqnarray}
with matrix $A$ defined as
\begin{equation}
A_{j,k} =  \left\{
\begin{array}{ll}
j(k+1) \qquad  & j\leq k  \\
A_{k,j} & j>k .
\end{array} \right.
\end{equation}
Writing this in favour of the occupation
numbers of antiparticles $m_j$ defined
in (\ref{antip}), we can simplify to
\begin{equation}
\e^{\displaystyle -\beta E_{\mbox{\scriptsize min}}}
= q^{\displaystyle \case{1}{4} \:  \vec{m}^T
C \:  \vec{m} },
\label{Emin}
\end{equation}
where $C$ denotes the Cartan matrix
of the simply-laced Lie algebra A$_{r-3}$.

As before, all other states with the same content
$n_1,\ldots,n_{r-2}$ are obtained from the minimal configuration
by moving particles to the left following M1-M6.
The combinatorics of this rearranging of particles has been
considered in section~\ref{secCG}. What has not yet been worked out
is the increase of energy
associated with the leftward motion of a particle.
Since we move one particle at a time,
we only have to calculate the energy increase
of the movement of a particle of charge $k$, relative
to a charge complex with charge $j>k$
positioned immediately to the left of this
particle.
Possibly the simplest way to go about is to first view
the complex and the particle as one single complex with
charge $j+k$. Suppose that relative to the origin
of this complex we have (local) minima and maxima of the
contour at
$0<x_1<y_1<x_2<y_2< \ldots < x_{n-1}<y_{n-1}<x_n<2k+2j$, labelling
minima by $y$ and maxima by $x$.
Note that the simplest possible scenario; the complex of charge
$j$ consisting of a single particle, corresponds to $n=2$.
The total weight of the complex relative to its own origin
is thus
\begin{equation}
\sum_{
\begin{array}{c}
\ell=1 \\
\ell\neq x_i,y_i
\end{array}}^{2j+2k-1} q^{k/2}.
\end{equation}
Now start moving the particle with charge $k$ one step to the left,
like shown in figure \ref{fig2}.
This results in a decrease of both $y_{n-1}$ and $x_n$ by 1.
The weight increase of this step is hence a factor
\begin{equation}
\frac{q^{(y_{n-1}+x_n+2)/2}}{q^{(y_{n-1}+x_n)/2}} = q.
\end{equation}
Taking another step to the left, again leads to a decrease
of $y_{n-1}$ and $x_n$ by 1 and we pick up another factor
$q$. This process can be repeated until $y_{n-1}-x_{n-1}=
x_n-y_{n-1}$, corresponding to the situation that the
maxima at $x_n$ and $x_{n-1}$ have equal height
(see figure~\ref{fig2}c).
{}From then on moving the particle a single step leads
to a decrease of $x_{n-1}$ and $y_{n-1}$ by 1.
However, this still leads to an increase by a single
factor $q$, and we are led to conclude that
each step of particle $k$ trough the complex yields
an extra factor $q$.

With this result we can simply carry out the $q$-analogue
of the previous subsection.
To do so we repeatedly
make use of the identity \cite{Andrews}
\begin{equation}
\sum_{k_1=0}^N \sum_{k_2=0}^{k_1} \ldots \sum_{k_M=0}^{k_{M-1}}
q^{k_1+k_2+\ldots+k_M} = \Mult{N+M}{M}{q}.
\label{NM}
\end{equation}
First we start moving the particles of charge $r-3$
by (\ref{NM}) this gives a Boltzmann factor
\begin{equation}
\sum_{k_1=0}^{2n_{r-2}} \sum_{k_2=0}^{k_1} \ldots
\sum_{k_{n_{r-3}}=0}^{k_{n_{r-3}-1}} q^{k_1+k_2+\ldots+k_{n_{r-3}}} =
\Mult{2 n_{r-2} + n_{r-3}}{n_{r-3}}{q} =
\Mult{n_{r-3} + m_{r-3}}{m_{r-3}}{q}.
\label{rmin2}
\end{equation}
Similarly, moving the particles with charge $r-4$ to the
left results in
\begin{equation}
\sum_{k_1=0}^{4n_{r-2}+2n_{r-3}} \sum_{k_2=0}^{k_1} \ldots
\sum_{k_{n_{r-4}}=0}^{k_{n_{r-4}-1}} q^{k_1+k_2+\ldots + k_{n_{r-4}}} =
\Mult{4 n_{r-2} + 2 n_{r-3} + n_{r-4}}{n_{r-4}}{q} =
\Mult{n_{r-4} + m_{r-4}}{m_{r-4}}{q}.
\end{equation}
In the general case of particles with charge $j$ we end up with
the contribution
\begin{eqnarray}
\lefteqn{
\sum_{k_1=0}^{2(r-2-j)n_{r-2}+ \ldots + 4n_{j+2}+2n_{j+1}}
\sum_{k_2=0}^{k_1} \ldots
\sum_{k_{n_j}=0}^{k_{n_j-1}} q^{k_1+k_2+\ldots+ k_{n_j}}  } \nonumber \\
& & \nonumber \\
&&  \qquad  \qquad =
\Mult{2(r-2-j)n_{r-2} + \ldots + 4 n_{j+2} + 2n_{j+1}+n_j}{n_j}{q} =
\Mult{n_j + m_j}{m_j}{q}.
\label{genjq}
\end{eqnarray}

{}From the results (\ref{rmin2})-(\ref{genjq}) and (\ref{Emin})
we obtain an expression for the partition function $Z$ as
\begin{equation}
Z(n_1,n_2,\ldots,n_{r-2};q)=
q^{\textstyle \frac{1}{4} \:  \vec{m}^T
C \: \vec{m} }
\prod_{j=1}^{r-3}
\Mult{m_j+n_j}{m_j}{q},
\end{equation}
with equations (\ref{antip}) and (\ref{constraint})
relating the antiparticle numbers $m_j$
to the particle numbers $n_j$.
We can finally end this section by stating the result for the
grand-canonical partition function,
\begin{equation}
\Xi_m(q) =
\sum_{\mbox{\scriptsize particle content}}
Z(n_1,n_2,\ldots,n_{r-2};q) =
\left. \sum \right.^{'} \;
q^{\textstyle \frac{1}{4} \:  \vec{m}^T
C \: \vec{m} }
\prod_{j=1}^{r-3}
\Mult{m_j+n_j}{m_j}{q},
\label{gcpf}
\end{equation}
where, as in (\ref{Fm}),
the second sum is over all solutions to (\ref{constraint})
with each entry of $\vec{m}$ being even.

\subsection{Mapping of configuration sums onto Fermi gas partition
function}\label{map}
\begin{figure}[hbt]
\centerline{\epsffile{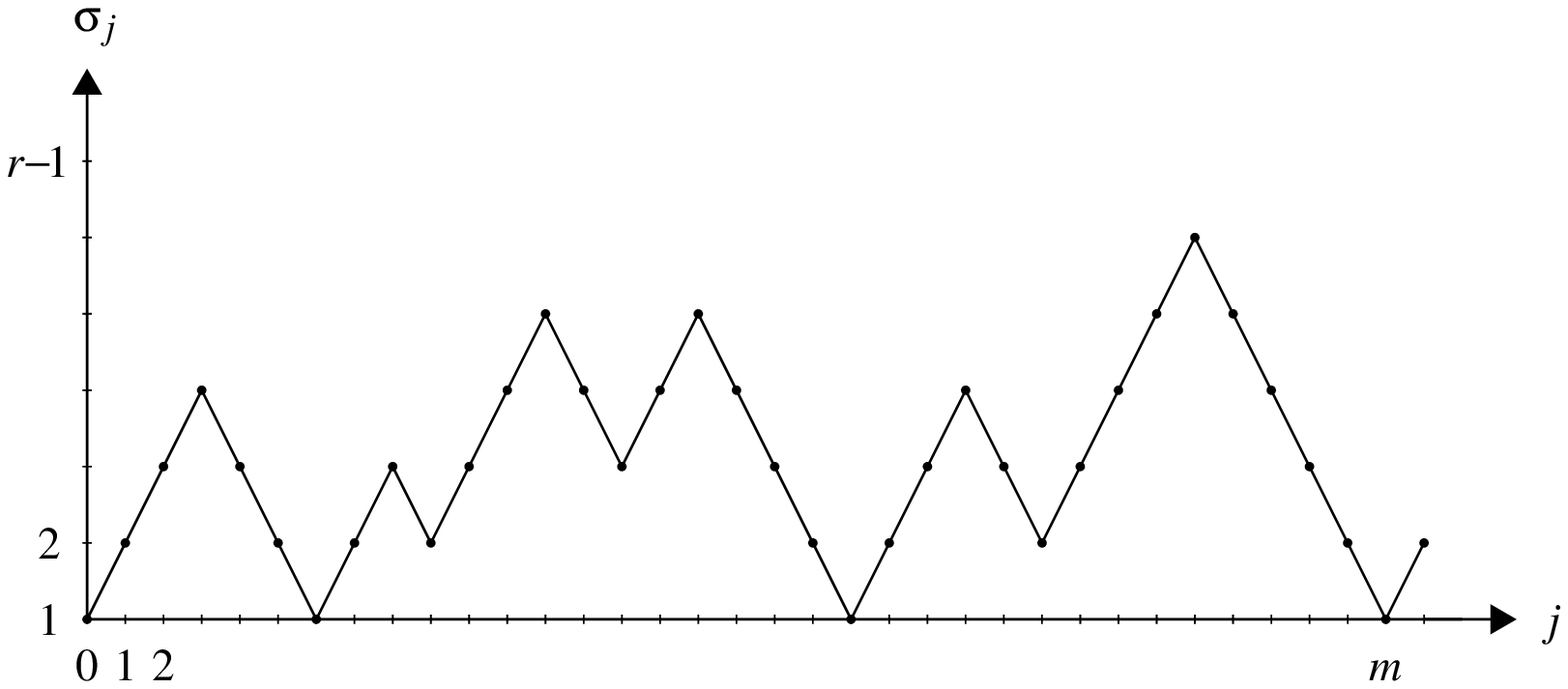}}
\caption{A contour of a typical admissible
sequence of spins $(\sigma_0,\sigma_1,\ldots,\sigma_{m+1})$.}
\label{fig7}
\end{figure}

To show that the result (\ref{gcpf}) is indeed the evaluation
of (\ref{confsums}) with $a=b=1$ and $c=2$, we  fix the
spins $\sigma_j\in \{1,2,\ldots,r-1\}$
such that $|\sigma_{j+1}-\sigma_j|=1$ for all $j$, and such that
$\sigma_0=\sigma_m=1, \; \sigma_{m+1}=2$. We call such a sequence
$(\sigma_0,\sigma_1,\ldots,\sigma_{m+1})$ {\em admissible}.
Next we plot $\sigma_j$ as a function of $j$ and
interpolate between $(j,\sigma_j)$ and $(j+1,\sigma_{j+1})$
by a straight line. We call the graph thus obtained the
contour of $(\sigma_0,\ldots,\sigma_{m+1})$.
A typical contour of an admissible sequence of sigma's
is shown in figure~\ref{fig7}.
{}From the definition (\ref{confsums}) of the configuration sums
we see that a straight line segment through $(j,\sigma_j)$,
corresponding to $|\sigma_{j-1}-\sigma_{j+1}|=2$, yields
a factor $q^{j/2}$.
The total weight of an admissible sequence can therefore
be written as
\begin{equation}
\e^{\displaystyle -\beta \sum_{j=1}^{m-1} E(j)}
\end{equation}
where the energy function $E(j)$ is precisely that
of the Fermi gas given by equation (\ref{energy}).
Since summing over all admissible sequences corresponds
to summing over all contours, which in turn corresponds to summing
over all charge configuration, we have established the
desired equivalence.

We note that particle-like interpretations
of admissible contours (or paths) have been previously formulated
for the minimal conformal series M$(2,k)$ in
refs.~[31-33].

The above equivalence also explains the defining rules R1 and R3.
{}From (\ref{confsums}) we have to count
each admissible sequence of spins just once.
However, interchanging two identical particles leads to
the same contour, and would therefore lead to multiple
counting of one and the same admissible sequence. Similarly
we have to impose R3 to avoid overcounting of admissible
sequences.

At the same time the mapping explains why the particular
choice of $X_m(1,1,2)$ is simplest to evaluate using
the fermionic technique.
If for e.g., we take  $a>1$, we have to introduce a
particle,
which, in separated unbound form, would have a profile of
a single straight line going down. This boundary particle
has a (fractional) charge of $(a-1)/2$. Similarly,
when $b=c-1>1$ we need a boundary particle of charge
$b/2$ with profile of a straight line going up.
These particles are in contrast
with the particles introduced so far,
which have origin and endpoint of equal height.
Since the motion of a (bulk)particle
through a boundary particle is very different
from that defined in R2 on page~\pageref{pR2},
the general $X_m(a,b,c)$ case becomes
quite complicated and technical.
For this reason
the computation of all one-dimensional configuration
sums will be deferred to part II.

\nsection{Polynomial and Rogers--Ramanujan type identities}\label{sec4}
Since the bosonic and the fermionic method to evaluate the configuration sum
$X_m(1,1,2)$
yield rather different type of expressions, we can of course combine
results to find some nontrivial identities.
Setting $a=b=1$ and $c=2$ in (\ref{the1}),
substituting (\ref{constraint}) into  (\ref{gcpf}), and
replacing $m$ by $2m$, yields
\begin{eqnarray}
\lefteqn{
\sum_{\begin{array}{c}
\scriptstyle m_1,\ldots,m_{r-3}\geq 0 \\
\scriptstyle \vec{m} \in 2 \Is^{r-3}
\end{array}}
q^{\case{1}{4} \: \textstyle
\vec{m}^T C \: \vec{m} }
\prod_{j=1}^{r-3}
\Mult{\case{1}{2}({\cal I} \: \vec{m}
)_j+m \delta_{j,1}}{m_j}{q} }
\nonumber \\ & & \nonumber \\
&&\qquad =
\sum_{j=-\infty}^{\infty} \left\{
q^{r(r-1)j^2+j}
\Mult{2m}{m-rj}{q}
-q^{r(r-1)j^2+(2r-1)j+1}
\Mult{2m}{m-1-rj}{q} \right\}.
\label{Polyid}
\end{eqnarray}
These identities are precisely the polynomial identities
for $X_{2m}(1,1,2)$
as conjectured by Melzer in ref.~\cite{Melzer1}.
A different proof of (\ref{Polyid})
has been published
by Berkovich~\cite{Berkovich}. In his proof Berkovich
has shown that a larger class of fermionic expressions,
corresponding to all $X_m(1,b,b\pm 1)$ solves the same recurrences
(\ref{recrel})
as the bosonic expressions (\ref{the1}) of ABF. To establish this,
elegant generalizations of the decompositions (\ref{qdecomp})
for the $q$-binomials were used.
One of the great merits of the recursive proof of ref.~\cite{Berkovich}
is that it proves a whole set of fermionic expressions at the same
time, and that complications with the boundary for general
$X_m(a,b,c)$ are avoided~\cite{Berkovich2}.
On the other hand,
the proof as given in this paper has the advantage of providing
detailed information relating BA solutions (see next section),
admissible sequences of spins occurring in the CTM calculation,
charge configurations of the Fermi gas and actual terms
within the sum on the left-hand side of (\ref{Polyid}).

In taking the $m\to\infty$ limit of (\ref{Polyid}),
using $\lim_{m\to\infty} \Mults{m}{k}{q} = 1/(q)_k$ and
$\lim_{m\to\infty} \Mults{2m}{m}{q} = 1/(q)_{\infty}$,
the following Rogers--Ramanujan type identities for
the $X_{1,1}^{(r-1,r)}(q)$ Virasoro characters arise:
\begin{eqnarray}
\lefteqn{
\sum_{\begin{array}{c}
\scriptstyle m_1,\ldots,m_{r-3}\geq m_{r-2}\equiv 0 \\
\scriptstyle \vec{m} \in 2 \Is^{r-3}
\end{array}}
\frac{q^{\case{1}{4} \: \textstyle
\vec{m}^T C \: \vec{m} }}{(q)_{m_1}}
\prod_{j=2}^{r-3}
\Mult{\case{1}{2}(m_{j-1}+m_{j+1})}{m_j}{q} }
\nonumber \\
&& \qquad \qquad \qquad \qquad \qquad =  \frac{1}{(q)_{\infty}}
\sum_{j=-\infty}^{\infty} \left\{
q^{r(r-1)j^2+j} -q^{r(r-1)j^2+(2r-1)j+1} \right\}.
\label{RR}
\end{eqnarray}
These identities
were first conjectured by the Stony Brook
group in ref.~\cite{KKMM2}.

To obtain Rogers--Ramanujan type identities for the branching
functions of the $\; \Integer_{r-2}$ parafermion conformal field
theories of ref.~\cite{ZF}, we
replace $q$ by $1/q$ in (\ref{Polyid})
and again let $m\to \infty$.
This results in
\begin{eqnarray}
\lefteqn{
\sum_{\begin{array}{c}
\scriptstyle m_1,\ldots,m_{r-3}\geq 0 \\
\scriptstyle ( C^{-1} \vec{m} )_{r-3} \in \Is
\end{array}}
\frac{q^{ \textstyle
\vec{m}^T C^{-1} \: \vec{m} }}{(q)_{m_1} \ldots (q)_{m_{r-3}}} }
\nonumber \\
&& \qquad \qquad \qquad
=  \frac{
q^{-\case{(r-3)(r-2)}{24r}}}{[(q)_{\infty}]^{r-3}}
\sum_{\alpha\in Q} \sum_{w\in W} \mbox{sgn}(w)\;
q^{\displaystyle \case{1}{2}r(r-1)\left|\alpha-\frac{r \rho -(r-1)
w(\rho)}{r(r-1)}\right|^2}\; ,
\label{paraf}
\end{eqnarray}
with $Q$ the root lattice, $W$ the Weyl group and $\rho$ the Weyl vector
of A$_{r-3}$. We note that the left-hand side of these identities are
the expressions for the branching functions obtained by
Lepowsky and Primc \cite{LP}.
To derive the above we have used the
level-rank equivalence between the $(r-1)$-state ABF
model in regime III and the level-2 A$_{r-3}$ Jimbo-Miwa-Okado model
in regime II \cite{JMO},
to rewrite the SU(2) type right-hand side of (\ref{Polyid})
into an SU($r-2$) type form~\cite{WP2} (see also the discussion
in ref.~\cite{BM}). For $r=4$ we note that
(\ref{RR}) and (\ref{paraf}) coincide.

\nsection{CTM versus TBA}\label{sec5}
The reader familiar with the TBA computations for the
ABF model of Bazhanov and Reshetikhin (BR)~\cite{BR1},
will have noticed the
many similarities between the TBA and the Fermi gas.
This section serves to indeed show
the mathematical equivalence between the fermionic
CTM calculations of section~\ref{sec3} and the TBA
calculations of ref.~\cite{BR1}.

First let us recall the Bethe Ansatz equations for the ABF
model \cite{BR1,BR2}
\begin{equation}
\omega^2
\left(\frac{\te(\frac{1}{2}\la + \ib \alpha_j)}{
\te(\frac{1}{2}\la - \ib \alpha_j)} \right)^m =
- \prod_{k=1}^{m/2} \frac{\te(\ib\alpha_j-\ib\alpha_k+ \lambda)}{
\te(\ib\alpha_j-\ib\alpha_k- \lambda)}, \qquad j=1,\ldots,\case{1}{2}m.
\end{equation}
Here $\omega=\exp(\ib \ell \la),$ $\ell=1,\ldots,r-1$,
$\la=\pi/r$ and each  set of roots
$\{\alpha_1,\ldots,\alpha_{m/2}\}$
yields an eigenvalue $\Lambda(u)$ of the row-to-row
transfer matrix.

Based on exact information for the cases $r=3$ (trivial 2-state
model)  and $r=4$ (Ising model), and
on a numerical investigation for other $r$ values,
BR formulated the following string hypothesis:
For $m\to \infty$ all solutions to the BAE consist of sets of strings,
with the allowed string types given by
\begin{equation}
\alpha_{\mu}^{(j)} = \alpha^{(j)} -
\frac{\ib \pi}{2 r}
(j+1-2\mu)
\qquad \mu=1,\ldots,j; \qquad j=1,\ldots,r-2,
\label{stringhypo}
\end{equation}
where
$\mu$ labels the $j$ roots within a string and
$j$ the different types of strings.
The real variable $\alpha^{(j)}$ is the string center.

As described in the introduction, we define our finitized model
by assuming the string hypothesis to be correct for finite $m$.
Let $n_j$ denote the number of strings of type $j$.
Then, since the total number of roots in a solution
equals $\case{1}{2} m$, we have the completeness relation
\begin{equation}
\sum_{j=1}^{r-2} j \, n_j = \case{1}{2} m.
\end{equation}
We note this is exactly the completeness relation (\ref{completeness})
for the particle numbers of the Fermi gas.

Substituting the hypothesis into the finite $m$ BAE, a set of
reduced equations is obtained which only involve the
string centers.
Carrying out the standard TBA procedures \cite{BR1,BGS,Dasmahapatra2},
the reduced equations yield the following set of constraint equations
for the number of strings $n_j$ and the number of {\em holes} $m_j$
(counting the missing Takahashi numbers):
\begin{equation}
m \, b_j = m_j + \sum_{k=1}^{r-2} B_{j,k} \, n_k,
\label{B}
\end{equation}
with
\begin{equation}
B_{j,k} =  \left\{
\begin{array}{ll}
\frac{\displaystyle 2(r-j)k}{\displaystyle r} \qquad  & j\geq k  \\
B_{k,j} & j<k
\end{array} \right. ,  \qquad \qquad b_j = \frac{r-j}{r} .
\end{equation}
Following BR we set $j=r-2$ in (\ref{B}) to find $m_{r-2}=0$.
Now eliminating $n_{r-2}$ yields
precisely the constraint equation (\ref{constraint}), and
we conclude that we can identify
the strings (holes) of type $j$ with the
particles (antiparticles) of charge $j$ $(-j)$ of the Fermi gas.
In other words, with each solution to the BAE for the finitized model
corresponds a configuration of the Fermi gas, and vice versa.
We remark here that
we can in fact easily establish a bijection between Bethe Ansatz
solutions and Coulomb gas configurations. So, for e.g.,
minimal configurations in the Fermi gas correspond to
solutions  without holes, and the leftward motion of a
particle with charge $j$  corresponds to shifting the Takahashi
numbers $I^{(j)}$ by ``inserting'' numbers $\tilde{I}^{(j)}$
corresponding to the creation of holes. Using the map from
the Fermi gas configurations to the configuration sums,
we have in fact a bijection between admissible sequences of spins
$(\sigma_1,\ldots,\sigma_{m+1})$ and solutions to the Bethe
Ansatz equations labelled by their Takahashi numbers.
In this context we should mention that several authors
\cite{BGS,Dasmahapatra1,Berkovich,Dasmahapatra2} have
indeed conjectured and/or
computed the total number of solutions to the Bethe
equations to be expression (\ref{Fm}), here counting the
number of charge configurations in the Fermi gas.

To further obtain the equations of TBA, we first let
$m \to \infty$ in the Fermi gas partition function
(\ref{gcpf}), and then take the limit $q\to 1^{-}$.
This procedure, developed in refs.~\cite{RS,NRT},
is equivalent to computing the $T\to 0$ asymptotics
of the entropy in TBA.
Writing
$ \lim_{m\to \infty} \Xi_m (q) = \sum_{k} a_k q^k$,
we have the asymptotics
\begin{equation}
\log a_k \sim \sqrt{3 \pi \left. \frac{S(T)}{T} \right|_{T=0}} \quad .
\end{equation}
Computing $a_k$ by steepest descend gives us \cite{DKKMM}
\begin{equation}
S(T) = -\frac{8\, T}{\pi} \sum_{j=1}^{r-3} [
L(\xi_j) - L(\eta_j) ] + o(T),
\end{equation}
with $L$ the Rogers dilogarithm function \cite{Lewin}
\begin{equation}
L(z) = \frac{1}{2} \int_0^z \left[\frac{\log(1-\zeta)}{\zeta}
+ \frac{\log\zeta}{1-\zeta} \right] \d{\zeta} .
\end{equation}
The numbers $\xi_j$ follow from the {\em TBA equations} \cite{BR1},
\begin{equation}
2 \log (1-\xi_j ) =
\sum_{k=1}^{r-3} C_{j,k} \, \log (\xi_k ),
\end{equation}
which arise here as the conditions for the saddle point.
The $\eta_{j+1}$ follow from the TBA equations
with $r$ replaced by $r-1$, and $\eta_1=1$.

Using the identity \cite{KR}
\begin{equation}
\sum_{j=1}^{r-3} [L(\xi_j) - L(\eta_j)] = -\frac{\pi^2}{6}
\left(1-\frac{6}{r(r-1)} \right),
\end{equation}
we finally get
\begin{equation}
\left. \frac{S(T)}{T} \right|_{T=0} = \frac{4 \pi}{3}
\left(1-\frac{6}{r(r-1)} \right),
\end{equation}
in accordance with the TBA result of BR.
{}From this last result the central charge of the ABF models
can be read off as \cite{Affleck}
\begin{equation}
c=1-\frac{6}{r(r-1)}.
\end{equation}

To conclude this section, we wish to make a comment on the
TBA calculations of BR as carried out in ref.~\cite{BR1}.
One of the shortcomings of their calculation is the
failure to also derive results for the scaling dimensions
of the ABF model.
However, from the equivalence between the Fermi gas and the TBA approach,
it is clear that the constraint equations (\ref{constraint})
which follow from the string hypothesis (\ref{stringhypo}) only describe
the vacuum or groundstate sector of the model. As argued before,
to  compute other configuration sums than $X_m(1,1,2)$,
(which corresponds to computing scaling dimensions)
additional boundary particles have to be introduced on top of the
Fermi gas. These extra particles result in a modification
of the constraint equations (\ref{constraint}).\footnote{These
modifications correspond to the ``additional inhomogeneities'' discussed in
refs.~\cite{Berkovich,BM}.}
We therefore believe that the string hypothesis as formulated by BR
is incomplete and does not describe all solutions to the
BAE. Indeed, the total number of eigenvalues of the
row-to-row transfer matrix of the ABF model
exceeds the number of configurations of the Fermi gas.
The fractional structure of the additional boundary particles needed
for the general configuration sums, seems to suggest the
existence of half-strings, which have roots in the
upper or lower part of the complex plane only.\footnote{
In a recent preprint \cite{BLZ} the generalization of TBA to compute
scaling dimensions was also briefly mentioned. How it relates
to the half-strings proposed here remains unclear at present.}

\nsection{Summary and discussion}\label{sec6}
In this paper we have presented a new method
for computing the one-dimensional configuration sums of the ABF model
in regime III.
Our approach, based on the interpretation of the configuration
sums as the grand-canonical partition function of a one-dimensional
gas of charged fermions,
results in polynomial expressions of fermionic type.
This opposed to the bosonic polynomials obtained by employing the
recurrence method of Andrews, Baxter and Forrester \cite{ABF}.
Combining both bosonic and fermionic techniques leads to a proof
of Melzer's polynomial identities \cite{Melzer1}, different
from the recursive proof by Berkovich \cite{Berkovich}.
In taking the thermodynamic limit, this also proofs
the Rogers--Ramanujan identities for unitary minimal Virasoro characters
as conjectured by Kedem {\em et al.} \cite{KKMM2}.
Since our fermionic method is mathematically equivalent to the
thermodynamic Bethe Ansatz calculations
for the ABF model of Bazhanov and Reshetikhin \cite{BR1}, we
have established a unification of the corner transfer matrix and
of the TBA technique. Here it is to be noted that in our calculations
the particle content of the model follows without making the usual
string hypothesis characterizing TBA calculations.

Although we have applied our method to the simplest possible
configuration sum of the ABF model, the other sums can be computed
be introducing additional boundary particles to the Fermi gas.
The details of this procedure will be treated in part II of this
paper \cite{W}.
What is more, our method is by no means restricted to the ABF model,
and we believe that all restricted solid-on-solid
models for which the TBA program has been
carried out \cite{BR1,BR2,BR3,Kuniba,BNW} admit a fermionic
computation of their configuration sums. We thus hope that our approach
leads to proofs of the many polynomial and Rogers--Ramanujan identities
conjectured in the recent literature \cite{KKMM1,KKMM2,WP2,Melzer3}.

We also note that our Fermi gas formulation of the ABF
configuration sums can be reformulated to obtain a partition
theoretic proof of Melzer's identities. Such a proof corresponds
to establishing a bijection between partitions with
prescribed hook differences \cite{ABBBFV} and charge
configurations of the Fermi gas. Work on this
bijection is currently in progress in collaboration with
O.~Foda~\cite{FW}.

Finally, it would be extremely
interesting to connect the approach to fermionic
character representations of Bouwknegt {\em et al.} [49-51]
using Yangian symmetries,
and of Georgiev using vertex operators \cite{Georgiev},
to that of this paper.

\section*{Acknowledgements}
I greatfully acknowledge the many hours Omar Foda
has spent in explaining me his partition theoretic insights
into Rogers--Ramanujan identities. His help has been invaluable.
I also wish to thank Alexander Berkovich and Barry McCoy
for their numerous suggestions to improve this manuscript.
I thank Peter Forrester, Bernard Nienhuis,
Aleks Owczarek and Paul Pearce for discussions and/or kind interest
in this work and David O'Brien for helping me out with {\em Mathematica}.
This work is supported by the Australian Research Council.

\newpage

\vspace*{5mm}

\vspace*{5mm}

\vspace*{5mm}

\begin{thebibliography}{99}

\bibitem{YY}
C.~N.~Yang and C.~P.~Yang,
{\em J.\ Math.\ Phys.} {\bf 10}:1115 (1969).

\bibitem{Takahashi}
M.~Takahashi,
{\em Prog.\ Theor.\ Phys.} {\bf 46}:401 (1971).

\bibitem{TS}
M.~Takahashi and M.~Suzuki,
{\em Prog.\ Theor.\ Phys.} {\bf 48}:2187 (1972).

\bibitem{BR1}
V.~V.~Bazhanov and N.~Yu~Reshetikhin,
{\em Int.\ J.\ Mod.\ Phys.\ A} {\bf 4}:115 (1989).

\bibitem{BR2}
V.~V.~Bazhanov and N.~Yu~Reshetikhin,
{\em J.\ Phys.\ A: Math.\ Gen.} {\bf 23}:1477 (1990).

\bibitem{BR3}
V.~V.~Bazhanov and N.~Yu~Reshetikhin,
{\em Prog.\ Theor.\ Phys.} {\bf 102}:301 (1990).

\bibitem{Baxter1}
R.~J.~Baxter,
{\em Exactly solved models in statistical mechanics}
(Academic Press, London, 1982).

\bibitem{ADM}
G.~Albertini, S.~Dasmahapatra and B.~M.~McCoy,
{\em Int.\ J.\ Mod.\ Phys.\ A} {\bf 7}, Suppl. 1A:1 (1992).

\bibitem{ABF}
G.~E.~Andrews, R.~J.~Baxter and P.~J.~Forrester,
{\em J.\ Stat.\ Phys.} {\bf 35}:193 (1984).

\bibitem{Melzer1}
E.~Melzer,
{\em Int.\ J.\ Mod.\ Phys.\ A} {\bf 9}:1115 (1994).

\bibitem{KKMM2}
R.~Kedem, T.~R.~Klassen, B.~M.~McCoy and E.~Melzer,
{\em Phys.\ Lett.} {\bf 304B}:263 (1993).

\bibitem{Berkovich}
A.~Berkovich,
{\em Nucl.\ Phys. B} {\bf 431}:315 (1994).

\bibitem{KM}
R.~Kedem and B.~M.~McCoy,
{\em J.\ Stat.\ Phys.} {\bf 71}:883 (1993).

\bibitem{DKMM}
S.~Dasmahapatra, R.~Kedem, B.~M.~McCoy and E.~Melzer,
{\em J.\ Stat.\ Phys.} {\bf 74}:239 (1994).

\bibitem{DKKMM}
S.~Dasmahapatra, R.~Kedem, T.~R.~Klassen, B.~M.~McCoy and E.~Melzer,
{\em Int.\ J.\ Mod.\ Phys.\ B} {\bf 7}:3617 (1993).

\bibitem{KKMM1}
R.~Kedem, T.~R.~Klassen, B.~M.~McCoy and E.~Melzer,
{\em Phys.\ Lett.} {\bf 307B}:68 (1993).

\bibitem{JMO}
M.~Jimbo, T.~Miwa and M.~Okado,
{\em Nucl.\ Phys.\ B} {\bf 300 [FS22]}:74 (1988).

\bibitem{KNS}
A.~Kuniba, T.~Nakanishi and J.~Suzuki,
{\em Mod.\ Phys.\ Lett.\ A} {\bf 8}:1649 (1993).

\bibitem{Dasmahapatra1}
S.~Dasmahapatra,
{\em String hypothesis and characters of coset CFT's}
preprint ICTP IC/93/91, hep-th/9305024.

\bibitem{Melzer2}
E.~Melzer,
{\em Lett.\ Math.\ Phys.} {\bf 31}:233 (1994).

\bibitem{Dasmahapatra2}
S.~Dasmahapatra,
{\em On State Counting and Characters}
preprint CMPS 94-103, hep-th/9404116.

\bibitem{FQ}
O.~Foda and Y.-H.~Quano,
{\em Polynomial identities of the Rogers--Ramanujan type}
preprint University of Melbourne No. 25-94, hep-th/9407191.

\bibitem{WP1}
S.~O.~Warnaar and P.~A.~Pearce,
{\em J.\ Phys.\ A: Math.\ Gen.} {\bf 27}:L891 (1994).

\bibitem{WP2}
S.~O.~Warnaar and P.~A.~Pearce,
{\em A-D-E Polynomial and Rogers--Ramanujan Identities},
preprint University of Melbourne No. 41-94, hep-th/9411009.

\bibitem{BM}
A.~Berkovich and B.~M.~McCoy,
{\em Continued Fractions and Fermionic Representations for
Characters of $M(p,p')$ Minimal Models},
preprint BONN-TH-94-28, ITPSB 94-060, hep-th/9412030.

\bibitem{Melzer3}
E.~Melzer,
{\em Supersymmetric Analogs of the Gordon--Andrews Identities,
and related TBA Systems},
preprint TAUP 2211-94, hep-th/9412154.

\bibitem{Huse}
D.~A.~Huse,
{\em Phys.\ Rev.\ Lett.} {\bf 49}:1121 (1982).

\bibitem{Baxter2}
R.~J.~Baxter,
{\em J.\ Stat.\ Phys} {\bf 26}:427 (1981).

\bibitem{Andrews}
G.~E.~Andrews,
{\em The Theory of Partitions}
(Addison-Wesley, Reading, Massachusetts, 1976).

\bibitem{Rocha}
A.~Rocha-Caridi,
in {\em Vertex Operators in Mathematics and Physics},
eds.\ J.~Lepowsky, S.~Mandelstam and I.~M.~Singer
(Springer, Berlin, 1985).

\bibitem{Bressoud}
D.~Bressoud,
{\em Lecture Notes in Math.} {\bf 1395}:140 (1987).

\bibitem{KRV}
J.~Kellendonk, M.~R\"osgen and R.~Varnhagen,
{\em Int.\ J.\ Mod\ Phys.\ A} {\bf 9}:1009 (1994).

\bibitem{RV}
M.~R\"osgen and R.~Varnhagen,
{\em Steps towards Lattice Virasoro Algebras: su(1,1)}
preprint BONN-TH-94-23, hep-th/9501005.

\bibitem{Berkovich2}
A.~Berkovich, private communication.

\bibitem{ZF}
A.~B.~Zamolodchikov and V.~A.~Fateev,
{\em Sov.\ Phys.\ JETP} {\bf 62}:215 (1985).

\bibitem{LP}
J.~Lepowsky and M.~Primc,
{\em Structure of the standard modules for the affine Lie algebra
A$_1^{(1)}$} Contemporary Mathematics, {\bf 46} (AMS, Providence, 1985).

\bibitem{BGS}
A.~Berkovich, C.~Gomez and G.~Sierra,
{\em Nucl.\ Phys. B} {\bf 415}:681 (1994).

\bibitem{RS}
B.~Richmond and G.~Szekeres,
{\em J.\ Austral.\ Math.\ Soc.\ (A)} {\bf 31}:362 (1981).

\bibitem{NRT}
W.~Nahm, A.~Recknagel and M.~Terhoeven,
{\em Mod.\ Phys.\ Lett. A} {\bf 8}:1835 (1993).

\bibitem{Lewin}
L.~Lewin,
{\em Polylogarithms and Associated Functions}
(Elsevier, Amsterdam, 1981).

\bibitem{KR}
A.~N.~Kirillov and N.~{Yu}.~Reshetikhin,
{\em J.\ Sov.\ Math.} {\bf 52}:3156 (1990).

\bibitem{Affleck}
I.~Affleck,
{\em Phys.\ Rev.\ Lett.} {\bf 56}:746 (1986).

\bibitem{BLZ}
V.~V.~Bazhanov, S.~L.~Lukyanov, A.~B.~Zamolodchikov,
{\em Integrable Structure of Conformal Field Theory,
Quantum KdV Theory and Thermodynamic Bethe Ansatz},
preprint CLNS 94/1316, hep-th/9412229.

\bibitem{W}
S.~O.~Warnaar,
{\em Fermionic solution of the Andrews-Baxter-Forrester model II:
proof of Melzer's polynomial identities}, in preparation.

\bibitem{Kuniba}
A.~Kuniba,
{\em Nucl.\ Phys.\ B} {\bf 389}:209 (1993).

\bibitem{BNW}
V.~V.~Bazhanov, B.~Nienhuis and S.~O.~Warnaar,
{\em Phys.\ Lett.} {\bf 322B}:198 (1994).

\bibitem{ABBBFV}
G.~E.~Andrews, R.~J.~Baxter, D.~M.~Bressoud, W.~J.~Burge,
P.~J.~Forrester and G.~Viennot,
{\em Europ.\ J.\ Comb.} {\bf 8}:341 (1987).

\bibitem{FW}
O.~Foda and S.~O.~Warnaar,
{\em Bijective proof of Melzer's polynomial identities: the
$\chi_{1,1}^{(p,p+1)}$ case},
preprint University of Melbourne No. 03-95.

\bibitem{BLS1}
P.~Bouwknegt, A.~W.~W.~Ludwig and K.~Schoutens,
{\em Phys.\ Lett.} {\bf 338B}:448 (1994).

\bibitem{BLS2}
P.~Bouwknegt, A.~W.~W.~Ludwig and K.~Schoutens,
{\em Spinon basis for higher level SU(2) WZW models},
preprint USC-94/20, hep-th/9412108.

\bibitem{BLS3}
P.~Bouwknegt, A.~W.~W.~Ludwig and K.~Schoutens,
{\em Affine and Yangian symmetries in $SU(2)_1$
conformal field theory},
preprint USC-94/21, hep-th/9412199.

\bibitem{Georgiev}
G.~Georgiev,
{\em Combinatorial construction of modules for
infinite-dimensional Lie algebras, I. Principal
subspace},
preprint Rutgers University, hep-th/9412054.




\end{thebibliography}
\end{document}